\documentclass[fleqn,11pt]{article}

\usepackage{amsfonts,amsmath,amssymb}
\usepackage{latexsym}
\usepackage{graphicx}
\usepackage{multicol}

\usepackage{amsfonts,amssymb,cite}
\usepackage{graphicx}

\def\noi{\noindent}

\newcommand{\Title}[1]{\noi {{\Large\bf #1}}\\[1ex]}

\def\Aunames#1{\noi{\bf #1}}
\def\au#1{${}^{#1}$}
\def\Addresses#1{\medskip\noi \protect
	\begin{description}\itemsep -3pt {\it #1} \end{description}}
\def\adr#1#2{\item[${}^{#1}$]{\it #2}}

\newcommand{\Abstract}[1]{\vskip 2mm \begin{center}
        \parbox{16.4cm}{\small\noi #1} \end{center}\medskip}

\def\email#1#2{\footnotetext[#1]{e-mail: #2}\addtocounter{footnote}{1}}

\def\nqq{\hspace*{-2em}}

\usepackage{color}

\def\Jl#1#2{#1 {\bf #2},\ }

\def\ApJ#1 {\Jl{Astroph. J.}{#1}}
\def\CQG#1 {\Jl{Class. Quantum Grav.}{#1}}
\def\DAN#1 {\Jl{Dokl. AN SSSR}{#1}}
\def\GC#1 {\Jl{Grav. Cosmol.}{#1}}
\def\GRG#1 {\Jl{Gen. Rel. Grav.}{#1}}
\def\IJMPD#1 {\Jl{Int. J. Mod. Phys. D}{#1}}
\def\JETF#1 {\Jl{Zh. Eksp. Teor. Fiz.}{#1}}
\def\JETP#1 {\Jl{Sov. Phys. JETP}{#1}}
\def\JHEP#1 {\Jl{JHEP}{#1}}
\def\JMP#1 {\Jl{J. Math. Phys.}{#1}}
\def\NPB#1 {\Jl{Nucl. Phys. B}{#1}}
\def\NP#1 {\Jl{Nucl. Phys.}{#1}}
\def\PLA#1 {\Jl{Phys. Lett. A}{#1}}
\def\PLB#1 {\Jl{Phys. Lett. B}{#1}}
\def\PRD#1 {\Jl{Phys. Rev. D}{#1}}
\def\PRL#1 {\Jl{Phys. Rev. Lett.}{#1}}

\def\lal{&&\nqq {}}

\def\beq{\begin{equation}}
\def\eeq{\end{equation}}
\def\bear{\begin{eqnarray}}
\def\bearr{\begin{eqnarray} \lal}
\def\ear{\end{eqnarray}}
\def\earn{\nonumber \end{eqnarray}}

\begin{document}
\twocolumn[


\Title{Low-frequency gravitational waves coupled with electromagnetic waves in material media}

\Aunames{A.N. Morozov\au{a;1} and I.V. Fomin\au{a;2}}

\Addresses{\small
\adr a {Bauman Moscow State Technical University,
  Moscow, 105005, Russia}}


\Abstract{The influence of the low-frequency gravitational waves coupled with electromagnetic waves in material media on the test masses is investigated. The propagation of coupled gravitational waves in rarefied gases and cold magnetized plasma is considered. It has been shown that under specific conditions the amplitude of the coupled gravitational waves in a media reaches values of the order of the amplitude of transverse gravitational waves from external astrophysical sources. The specific properties of longitudinal gravitational waves coupled with electromagnetic waves  in a medium in relation to transverse gravitational waves from external sources are considered as well.}

] 
\email 1 {amor59@mail.ru}
\email 2 {ingvor@inbox.ru}

\section{Introduction}

Wave solutions of Einstein's equations describing gravitational waves (GWs) perturbations of the space-time metric have been actively studied in the context of the application of the General Relativity to the description of gravitational interaction~\cite{Maggiore:2018sht,Pustovoit:2016zyt,Rudenko:2017asr}.
Direct registration of the gravitational waves from the black holes and neutron stars mergers~\cite{LIGOScientific:2016aoc,LIGOScientific:2016sjg,LIGOScientific:2017zic} based on the interferometric method proposed in~\cite{Gertsenshtein:1962kfm} confirmed the physical validity of gravitational-wave solutions of Einstein's equations. The basis of this method of registration is the influence of gravitational waves on test masses (interferometer mirrors) and analysis of the corresponding change in intensity on the photodetector due to the phase shift induced by the periodic displacement of the interferometer's mirrors~\cite{Maggiore:2018sht,Gertsenshtein:1962kfm}.
Although many other types of the GW's detection methods are considered~\cite{Aggarwal:2020olq}, at the moment, only the interferometric method allows for the direct registration of gravitational waves~\cite{LIGOScientific:2016aoc,LIGOScientific:2016sjg,LIGOScientific:2017zic}.

Electromagnetic radiation itself is also a source of gravitational field~\cite{Bonnor:1969mfs, Aichelburg:1970dh,Grishchuk:1973qz, Bonnor:2009zza,vanHolten:2008ts,Ratzel:2015nqf}. Previously, the different methods for estimating the characteristics of gravitational waves induced by electromagnetic waves were considered.
For a powerful laser pulse interacted with a material medium, the maximal amplitude of the gravitational waves induced by the shock waves was estimated as $h_{0}\sim10^{-40}$~\cite{Kadlecova:2016fbx,Fomin:2020ovn}. Direct consideration of an electromagnetic wave as a source of gravitational waves without interaction with the medium leads to the following estimation of the maximal amplitude $h_{0}\sim10^{-36}-10^{-37}$~\cite{Lageyre:2021cxe,Atonga:2023psc}.
In order to obtain these estimates, the transverse-traceless (TT) gauge of gravitational waves perturbations of the space-time metric was considered~\cite{Kadlecova:2016fbx,Fomin:2020ovn,Lageyre:2021cxe,Atonga:2023psc}.
This approach is due to the fact that in a vacuum the longitudinal gravitational wave's components don't correspond to the real perturbations of the space-time metric, and transverse-traceless gauge is the relevant one~\cite{Maggiore:2018sht}. However, in the presence of gravitational sources it is not possible to define the full space-time metric perturbations in the TT-gauge~\cite{Flanagan:2005yc,Fanizza:2021ngq}.
Thus, the structure of the solutions for the gravitational waves coupled with the source is different from the case of the gravitational waves in a vacuum.

The various approaches to analyse the influence of the material fields, including plasma and magnetic fields on the characteristics of the gravitational waves was considered, for example, in~\cite{Galtsov:1983oog,Macedo:1983wcr, Servin:2003cf,Isliker:2006cq,Brodin:2009yy,Bamba:2018cup,Asenjo:2020aea,Moretti:2019yhs,VillarrealFasanelli:2024gmz}.
The coupled gravitational and electromagnetic waves in the empty space, material media and external magnetic field was also considered in~\cite{STFI2020,Morozov:2020eiu,Morozov:2021ork,Morozov:2026pns}. Within the framework of this approach, wave solutions of a weak gravitational field equations were considered, implying both longitudinal and transverse components of gravitational waves.
Also, in~\cite{Morozov:2026pns} it was shown that in cold magnetized plasma the energy flux density of gravitational waves coupled with an electromagnetic wave increases  compared to the case of vacuum due to the deviations of the refractive index from unity.

In this paper, we consider coupled gravitational waves for the case of deviation of electromagnetic radiation from a monochromatic one during propagation in material media with and without an external magnetic field. For a transparent analytical description of the gravitational-wave counterpart of electromagnetic radiation, we will consider the plane electromagnetic waves.
Based on the analysis of solutions of the gravitational field equations, it is shown that in this case, gravitational waves induced by an electromagnetic wave are also coupled with the material medium. In an analysis of the influence of the coupled gravitational waves on test masses, it was shown that they induce tidal forces similar to transverse gravitational waves from external sources.
The influence of the refractive index of the medium on the characteristics of coupled gravitational waves and geometric invariants of the space-time is also considered.

\section{Gravitational field equations}

In order to describe gravitational waves coupled with an electromagnetic waves, we consider the Einstein equations in the following form~\cite{Maggiore:2018sht}
\begin{align}
&R_{\mu\nu} = \frac{8\pi G}{c^{4}} \left(T_{\mu\nu} - \frac{1}{2} g_{\mu\nu} T\right),
\label{eq:Einstein}
\end{align}
where $G$ is the gravitational constant, $c$ is the speed of light in vacuum, $R_{\mu\nu} = g^{\sigma\rho} R_{\sigma\mu\rho\nu}$ is the Ricci tensor, $T_{\mu\nu}$ is the energy-momentum tensor of the electromagnetic field, $T = g^{\mu\nu} T_{\mu\nu}$, $\mu,\nu = 0,1,2,3$.

In the case of a weak gravitational field, the space-time metric can be defined as
\begin{align}
&g_{\mu\nu} = \eta_{\mu\nu} + h_{\mu\nu}, \quad |h_{\mu\nu}| \ll 1,
\label{eq:metric_pert}
\end{align}
where $\eta_{\mu\nu}$ is the metric tensor of the Minkowski space-time.

In equation (\ref{eq:Einstein}) we use the following notation
\begin{align}
&\tilde{T}_{\mu\nu} \equiv T_{\mu\nu} - \frac{1}{2} g_{\mu\nu} T \approx T_{\mu\nu} - \frac{1}{2} \eta_{\mu\nu} T,
\label{eq:T_tilde_def}
\end{align}
where the last condition means that we neglect the influence of a weak gravitational field on the electromagnetic field.

The Riemann tensor for the case of a weak gravitational field can be written as follows~\cite{Maggiore:2018sht}
\begin{align}
\nonumber
&R_{\sigma\alpha\rho\beta}=\frac{1}{2} \big(\partial_{\alpha} \partial_{\rho} h_{\sigma\beta} + \partial_{\sigma} \partial_{\beta} h_{\alpha\rho}\\
&- \partial_{\alpha} \partial_{\beta} h_{\sigma\rho} - \partial_{\sigma} \partial_{\rho} h_{\alpha\beta} \big),
\label{eq:Riemann_weak}
\end{align}
where $\partial_{\alpha} = \frac{\partial}{\partial x^{\alpha}}$.

Now, we consider the harmonic gauge conditions~\cite{Maggiore:2018sht}
\begin{align}
&\partial_{\mu} \left( \eta^{\mu\sigma} h_{\sigma\nu} - \frac{1}{2} \delta_{\nu}^{\mu} h \right) = 0,
\label{eq:harmonic_gauge}
\end{align}
where
\begin{align}
&h = \eta^{\mu\nu} h_{\mu\nu} = h_{00} - h_{11} - h_{22} - h_{33}.
\label{eq:h_trace}
\end{align}

Under the harmonic gauge conditions \eqref{eq:harmonic_gauge}, equations \eqref{eq:Einstein} reduce to the following form
\begin{align}
&-\eta^{\sigma\alpha} \partial_{\sigma} \partial_{\alpha} h_{\mu\nu} = \frac{16\pi G}{c^{4}} \tilde{T}_{\mu\nu}.
\label{eq:wave_eq}
\end{align}

In the case of gravitational wave propagation in vacuum without a source ($\tilde{T}_{\mu\nu}=0$), GWs contain only two degrees of freedom according to the transverse-traceless (TT) gauge. In the case under consideration, solutions of equations \eqref{eq:wave_eq} may differ from ones obtained in the TT-gauge.

\section{Energy-momentum tensor of electromagnetic waves}

Let us consider two plane electromagnetic waves with frequencies $\omega_{1}$ and $\omega_{2}$ propagating in a some material medium with dielectric permittivity $\varepsilon$ and magnetic permeability $\mu=1$ and transverse magnetic field $H_{00}$ in the direction of the axis $x^{1}=x$.
These electromagnetic waves can be described by the following expressions for the electric and magnetic field strengths
\begin{align}
\label{eq:Ey}
\nonumber
&E_{y} = E_{1} \cos \left( \omega_{1} \left( t - \frac{nx}{c} \right) \right)\\
&+ E_{2} \cos \left( \omega_{2} \left( t - \frac{nx}{c} \right) \right),\\
\nonumber
\label{eq:Hz}
&H_{z} = H_{1} \cos \left( \omega_{1} \left( t - \frac{nx}{c} \right) \right)\\
&+ H_{2} \cos \left( \omega_{2} \left( t - \frac{nx}{c} \right) \right) + H_{00},
\end{align}
where $n=\sqrt{\varepsilon}$ is the refractive index of the medium, $H_{00}$ is a constant magnetic field, which is transverse to the direction of electromagnetic wave propagation.

For simplicity, we will further assume that
\begin{align}
&E_{1} = E_{2} = E_{0},
\label{eq:E0_equal} \\
&H_{1} = H_{2} = H_{0},
\label{eq:H0_equal}
\end{align}
where the amplitudes of the electric $E_{0}$ and magnetic $H_{0}$ field strengths are related as follows
\begin{align}
&H_{0} = \sqrt{\varepsilon} E_{0} = n E_{0}.
\label{eq:H0_E0_rel}
\end{align}

The components of the corresponding energy-momentum tensor of electromagnetic field  for the cold magnetized plasma or rarefied gas with constant temperature $T_{0}$ can be written as follows~\cite{Morozov:2020eiu,Morozov:2026pns}
\begin{align}
\label{Energy}
&T_{00}=\frac{1}{8\pi}\left(\varepsilon E^{2}_{y}+H^{2}_{z}\right),\\
\label{Pointing}
&T_{01}=T_{10}=-\frac{1}{4\pi}\left(E_{y}H_{z}+E_{y}H_{00}\right),\\
\label{Maxvell}
\nonumber
&T_{aa}=-\frac{\varepsilon}{4\pi} E_{a}E_{a} - \frac{1}{4\pi} H_{a}H_{a}\\
&+ \frac{1}{8\pi}\left(\varepsilon - \rho\left(\frac{\partial\varepsilon}{\partial\rho}\right)_{T_{0}}\right)E^{2}_{y}\\
\nonumber
&+ \frac{1}{8\pi}\left(1 - \frac{\rho}{\varepsilon}\left(\frac{\partial\varepsilon}{\partial\rho}\right)_{T_{0}}\right)H^{2}_{z},
\end{align}
where $a=1,2,3$, and dielectric permittivity is
\begin{align}
\label{vareps}
&\varepsilon(\rho)=1+\kappa\rho,
\end{align}
where $\rho=mN$ is the density of electrons in plasma or the density of the rarefied gas.

The constant $\kappa$ for a cold magnetized plasma is~\cite{Morozov:2026pns}
\begin{eqnarray}
\label{A2}
  \kappa = \begin{cases}
    \frac{4\pi e^{2}c}{m^{2}\omega^{2}}, & \text{for $H_{00}=0$}, \\
    \frac{4\pi ec}{H_{00}m^{3/2} \delta\tilde{\omega}}, & \text{for $H_{00}\neq0$},
  \end{cases}
\end{eqnarray}
where $e$ and $m$ are the electron's charge and mass, and $\delta\tilde{\omega}$ is the difference between the frequency of electromagnetic wave $\omega$ and the plasma frequency $\omega_{0}$ for the condition $\frac{\delta\tilde{\omega}}{\omega_{0}}\ll1$.

For the case of a rarefied gas with $n\approx1$ the constant $\kappa$ in expression (\ref{vareps}) depends on its molecular weight and polarizability.

Thus, from expressions (\ref{eq:Ey})--(\ref{eq:Hz}) and (\ref{Energy})--(\ref{Maxvell}) we obtain
\begin{align}
\nonumber
&\tilde{T}_{00} = \frac{\left(3-n^{2}\right)E_{0}^{2}}{16\pi} u_{2n}(x,t)
		+ \frac{3-n^{2}}{16\pi n^{2}} H_{00}^{2} \\
&+ \frac{3-n^{2}}{8\pi n} E_{0} H_{00} u_{1n}(x,t),
\label{eq:T_tilde_00}
\end{align}
\begin{align}
&\tilde{T}_{01} = \tilde{T}_{10} = -\frac{n E_{0}^{2}}{8\pi} u_{2n}(x,t) \nonumber\\
&	- \frac{E_{0} H_{00}}{4\pi} u_{1n}(x,t),
\label{eq:T_tilde_01}
\end{align}
\begin{align}
\nonumber
&\tilde{T}_{11} = \frac{\left(3n^{2}-1\right)E_{0}^{2}}{16\pi} u_{2n}(x,t)
+\frac{5n^{2}-3}{16\pi n^{2}} H_{00}^{2} \\
&+ \frac{3n^{2}-1}{8\pi n} E_{0} H_{00} u_{1n}(x,t),
\label{eq:T_tilde_11}
\end{align}
\begin{align}
\nonumber
&\tilde{T}_{22} = \frac{\left(n^{2}-1\right)E_{0}^{2}}{16\pi} u_{2n}(x,t)
+ \frac{5n^{2}-3}{16\pi n^{2}} H_{00}^{2}\\
& + \frac{3n^{2}-1}{8\pi n} E_{0} H_{00} u_{1n}(x,t),		
\label{eq:T_tilde_22}
\end{align}
\begin{align}
\nonumber
&\tilde{T}_{33} = \frac{\left(n^{2}-1\right)E_{0}^{2}}{16\pi} u_{2n}(x,t)
- \frac{n^{2}+1}{16\pi n^{2}} H_{00}^{2} \\
&- \frac{n^{2}+1}{8\pi n} E_{0} H_{00} u_{1n}(x,t),
\label{eq:T_tilde_33}
\end{align}
where
\begin{align}
\nonumber
& u_{1n}(x,t) = \cos\left( \omega_{1}\left(t-\frac{nx}{c}\right) \right)\\
&		+ \cos\left( \omega_{2}\left(t-\frac{nx}{c}\right) \right),
\label{eq:u1n}
\end{align}
and
\begin{align}
\label{eq:u2n_compact}
\nonumber
&u_{2n}(x,t) = \left[ \cos\Phi_{1} + \cos\Phi_{2} \right]^{2}\\
&= 2 + \cos(2\Phi_{1}) + \cos(2\Phi_{2})\\
\nonumber
&+ 2\left[ \cos(\Phi_{1}+\Phi_{2}) + \cos(\Phi_{2}-\Phi_{1}) \right],
\end{align}
where $\Phi_{i} = \omega_{i}\left(t-\frac{nx}{c}\right)$.

Further, we consider the case when the frequencies $\omega_{1}$ and $\omega_{2}$ of the electromagnetic waves are close to each other
\begin{align}
&\omega_{2} = \omega_{1} + \delta\omega,
\label{eq:omega_close}
\end{align}
where $\delta\omega \ll \omega_{1,2}\simeq\omega_{0}$.

At the same time, we assume that the plasma resonance condition~\cite{Gershman_1957,Bellan_2006}
\begin{align}
&\frac{\omega_{0}}{\delta\omega} < n,
\label{eq:plasma_res_cond}
\end{align}
for waves with frequencies $\omega_{1}$ and $\omega_{2}$ is satisfied.

Under this condition, we obtain
\begin{align}
&u_{1n}(x,t) = 2\cos\left( \omega_{0}\left(t-\frac{nx}{c}\right) \right),
\label{eq:u1n_simp} \\
\nonumber
&u_{2n}(x,t) = 2 + 4\cos\left( 2\omega_{0}\left(t-\frac{nx}{c}\right) \right)\\
&+ 2\cos\left( \delta\omega\left(t-\frac{nx}{c}\right) \right).
\label{eq:u2n_simp}
\end{align}

Also, the tensor $\tilde{T}_{\mu\nu}$ can be considered as the sum of constant and variable components, with frequencies $\omega_{0}$, $2\omega_{0}$, and $\delta\omega$, namely
\begin{align}
&\tilde{T}_{\mu\nu} = \tilde{T}_{\mu\nu}^{(\text{const})} + \tilde{T}_{\mu\nu}^{(\omega_{0})}
		+ \tilde{T}_{\mu\nu}^{(2\omega_{0})} + \delta\tilde{T}_{\mu\nu}.
\label{eq:T_tilde_decomp}
\end{align}

Therefore, we can consider only the low-frequency component of gravitational-wave perturbations of the space-time metric.  We will also consider the solutions of equations (\ref{eq:wave_eq}) as the difference between the perturbations of the space-time metric in the medium and in a vacuum.

Thus, we will consider variable components of the tensor $\delta\tilde{T}_{\mu\nu}$ in the following form
\begin{align}
\label{eq:deltaT_00}
&\delta\tilde{T}_{00} = \frac{\left(3-n^{2}\right)E_{0}^{2}}{8\pi} \cos\left( \delta\omega\left(t-\frac{nx}{c}\right) \right) \nonumber\\
&-\frac{E_{0}^{2}}{4\pi} \cos\left( \delta\omega\left(t-\frac{x}{c}\right) \right),
\end{align}
\begin{align}
\label{eq:deltaT_01}
&\delta\tilde{T}_{01} = \delta\tilde{T}_{10} = -\frac{n E_{0}^{2}}{4\pi} \cos\left( \delta\omega\left(t-\frac{nx}{c}\right) \right)\nonumber\\
&+\frac{E_{0}^{2}}{4\pi} \cos\left( \delta\omega\left(t-\frac{x}{c}\right) \right),
\end{align}
\begin{align}
\label{eq:deltaT_11}
&\delta\tilde{T}_{11} = \frac{\left(3n^{2}-1\right)E_{0}^{2}}{8\pi} \cos\left( \delta\omega\left(t-\frac{nx}{c}\right) \right)\nonumber\\
&-\frac{E_{0}^{2}}{4\pi} \cos\left( \delta\omega\left(t-\frac{x}{c}\right) \right),
\end{align}
\begin{align}
\label{eq:deltaT_22}
&\delta\tilde{T}_{22} = \frac{\left(n^{2}-1\right)E_{0}^{2}}{8\pi} \cos\left( \delta\omega\left(t-\frac{nx}{c}\right) \right),
\end{align}
\begin{align}
&\delta\tilde{T}_{33} = \frac{\left(n^{2}-1\right)E_{0}^{2}}{8\pi} \cos\left( \delta\omega\left(t-\frac{nx}{c}\right) \right),
\label{eq:deltaT_33}
\end{align}

In order to describe the low-frequency gravitational waves with frequency $\delta\omega$ coupled to the electromagnetic wave with the components \eqref{eq:deltaT_00}--\eqref{eq:deltaT_33} of the tensor $\delta\tilde{T}_{\mu\nu}$, equations \eqref{eq:wave_eq} can be written in the following form
\begin{align}
&-\left(\frac{\partial^{2}}{c^{2}\partial t^{2}} - \frac{\partial^{2}}{\partial x^{2}} \right) h_{\mu\nu}
		= \frac{16\pi G}{c^{4}} \delta\tilde{T}_{\mu\nu}.
\label{eq:wave_eq_delta}
\end{align}

We also note that, based on the form of tensor $\tilde{T}_{\mu\nu}$ with components \eqref{eq:deltaT_00}--\eqref{eq:deltaT_33}, we can consider the conversion between gravitational and electromagnetic waves in an external magnetic field~\cite{Kolosnitsyn:2015zua} for the case of the coupled gravitational waves with frequency $\delta\omega$ as negligible effect.

\section{Solutions of gravitational field equations}

Solutions of equations \eqref{eq:wave_eq_delta} with the components of the tensor \eqref{eq:deltaT_00}--\eqref{eq:deltaT_33} can be written as follows:
\begin{align}
&h_{00} = \frac{2 G E_{0}^{2} (n^{2}-3)}{c^{2} (\delta\omega)^{2} (n^{2}-1)}
          \cos\left( \delta\omega\left(t-\frac{nx}{c}\right) \right) \nonumber\\
& - \frac{2 G E_{0}^{2} (n^{2}-3)}{c^{2} (\delta\omega)^{2} (n^{2}-1)}
          \cos\left( \delta\omega\left(t-\frac{x}{c}\right) \right)\nonumber\\
&+\frac{2 E_{0}^{2} G x}{\delta\omega c^{3}}\sin\left( \delta\omega\left(t-\frac{x}{c}\right) \right),
\label{eq:h_00}
\end{align}
\begin{align}
&h_{01} = h_{10} = \frac{4 n G E_{0}^{2}}{c^{2} (\delta\omega)^{2} (n^{2}-1)}
                   \cos\left( \delta\omega\left(t-\frac{nx}{c}\right) \right) \nonumber \\
& - \frac{4 n G E_{0}^{2}}{c^{2} (\delta\omega)^{2} (n^{2}-1)}
                   \cos\left( \delta\omega\left(t-\frac{x}{c}\right) \right)\nonumber\\
&-\frac{2 E_{0}^{2} G x}{\delta\omega c^{3}}\sin\left( \delta\omega\left(t-\frac{x}{c}\right) \right),
\label{eq:h_01}
\end{align}
\begin{align}
&h_{11} = -\frac{2 G E_{0}^{2} (3n^{2}-1)}{c^{2} (\delta\omega)^{2} (n^{2}-1)}
          \cos\left( \delta\omega\left(t-\frac{nx}{c}\right) \right)  \nonumber\\
& + \frac{2 G E_{0}^{2} (3n^{2}-1)}{c^{2} (\delta\omega)^{2} (n^{2}-1)}
          \cos\left( \delta\omega\left(t-\frac{x}{c}\right) \right)\nonumber\\
&+\frac{2 E_{0}^{2} G x}{\delta\omega c^{3}}\sin\left( \delta\omega\left(t-\frac{x}{c}\right) \right),
\label{eq:h_11}
\end{align}
\begin{align}
&h_{22} = h_{33} = -\frac{2 G E_{0}^{2}}{c^{2} (\delta\omega)^{2}}
                   \cos\left( \delta\omega\left(t-\frac{nx}{c}\right) \right) \nonumber \\
& + \frac{2 G E_{0}^{2}}{c^{2} (\delta\omega)^{2}}
                   \cos\left( \delta\omega\left(t-\frac{x}{c}\right) \right).
\label{eq:h_22_33}
\end{align}

In the general case, solutions \eqref{eq:h_00}--\eqref{eq:h_22_33} are determined up to gauge functions $h_{\mu\nu} \to h_{\mu\nu} + \xi_{\mu\nu}$, which are defined in such a way as to obtain regular solutions of equations \eqref{eq:wave_eq_delta} in the vacuum limit $n\to1$ and in the non-wave limit $\delta\omega\to0$ as well.

We also note that the solutions of coupled gravitational waves with frequencies $\omega_{0}$ and $2\omega_{0}$ were considered earlier in~\cite{Morozov:2026pns}. The amplitudes of these coupled gravitational waves $h_0\sim(\omega_{0})^{-2}$ and $h_0\sim(2\omega_{0})^{-2}$ are significantly smaller than ones in solutions \eqref{eq:h_00}--\eqref{eq:h_22_33}.

Now, we consider the components of the Riemann tensor corresponding to these solutions.

Based on expressions \eqref{eq:Riemann_weak} and \eqref{eq:h_00}--\eqref{eq:h_22_33} we obtain the following components of the Riemann tensor
\begin{align}
\nonumber
&R_{1010}  =\partial_{0} \partial_{1} h_{01}
 - \frac{1}{2} \left( \partial_{0} \partial_{0} h_{11}
		+ \partial_{1} \partial_{1} h_{00} \right) \nonumber \\
&= \frac{G E_{0}^{2} (n^{2}-1)}{c^{4}}
      \cos\left( \delta\omega\left(t-\frac{nx}{c}\right) \right) \nonumber \\
& +\frac{2 G E_{0}^{2} (n-1)}{c^{4}(n+1)}
      \cos\left( \delta\omega\left(t-\frac{x}{c}\right) \right),
\label{eq:R_0101}
\end{align}
\begin{align}
&R_{2020} = R_{3030} = -\frac{1}{2} \partial_{0} \partial_{0} h_{22}= -\frac{1}{2} \partial_{0} \partial_{0} h_{33} \nonumber \\
&= -\frac{G E_{0}^{2}}{c^{4}} \cos\left( \delta\omega\left(t-\frac{nx}{c}\right) \right)\nonumber\\
&+\frac{G E_{0}^{2}}{c^{4}} \cos\left( \delta\omega\left(t-\frac{x}{c}\right) \right),
\end{align}
\begin{align}
&R_{0313} = R_{0212} = -\frac{1}{2} \partial_{0} \partial_{1} h_{22}
                      = -\frac{1}{2} \partial_{0} \partial_{1} h_{33} \nonumber \\
&= \frac{n G E_{0}^{2}}{c^{4}} \cos\left( \delta\omega\left(t-\frac{nx}{c}\right) \right)\nonumber \\
&- \frac{G E_{0}^{2}}{c^{4}} \cos\left( \delta\omega\left(t-\frac{x}{c}\right) \right),
\label{eq:R_0313}
\end{align}
\begin{align}
&R_{1212} = R_{1313} = -\frac{1}{2} \partial_{1} \partial_{1} h_{22}
                      = -\frac{1}{2} \partial_{1} \partial_{1}h_{33} \nonumber \\
&= - \frac{n^{2} G E_{0}^{2}}{c^{4}} \cos\left( \delta\omega\left(t-\frac{nx}{c}\right) \right)\nonumber \\
&+\frac{G E_{0}^{2}}{c^{4}} \cos\left( \delta\omega\left(t-\frac{x}{c}\right) \right).
\label{eq:R_1212}
\end{align}

The remain non-zero components can be determined based on the properties of the Riemann tensor~\cite{Maggiore:2018sht}.

Solutions \eqref{eq:h_00}--\eqref{eq:h_22_33} in the limit $\delta\omega\to0$ can be written as follows
\begin{align}
\label{eq:energynw}
&h^{(nw)}_{00}=\frac{G E_0^{2}  \left(n^{2}-3\right) x\left(2 t c - n x - x\right)}{\left(n+1\right) c^{4}}\nonumber\\
&+\frac{2G E_0^{2} \left(t c - x\right)x}{c^{4}},\\
\label{eq:energynw2}
&h^{(nw)}_{01}=\frac{2G E_0^{2}  n x\left(2 t c - n x - x\right) }{\left(n+1\right) c^{4}}\nonumber\\
&-\frac{2 \left(t c - x\right) G E_0^{2} x}{c^{4}},\\
\label{eq:energynw3}
&h^{(nw)}_{11}=-\frac{G E_0^{2}\left(3 n^{2}-1\right) x\left(2 t c - n x - x\right) }{\left(n+1\right) c^{4}}\nonumber\\
&+\frac{2 \left(t c - x\right) G E_0^{2} x}{c^{4}},\\
\label{eq:energynw4}
&h^{(nw)}_{22}=-\frac{G E_0^{2}\left(n-1\right)x \left(2 t c - n x - x\right)}{c^{4}},\\
\label{eq:energynw5}
&h^{(nw)}_{33}=h^{(nw)}_{22}.
\end{align}

The components \eqref{eq:deltaT_00}--\eqref{eq:deltaT_33} in the limit $\delta\omega\to0$ are reduced to the following form
\begin{align}
\label{eq:deltaT_00nw}
&\delta\tilde{T}_{00}^{(nw)} = \delta\tilde{T}_{01}^{(nw)}=\delta\tilde{T}_{10}^{(nw)}=-\delta\tilde{T}_{22}^{(nw)}\nonumber\\
&=-\delta\tilde{T}_{33}^{(nw)}=-\frac{(n^{2}-1)E^{2}_{0}}{8\pi},\\
\label{eq:deltaT_11nw}
&\delta\tilde{T}_{11}^{(nw)} = \frac{3(n^{2}-1)E^{2}_{0}}{8\pi}.
\end{align}

It should be noted that the components of tensor $\delta\tilde{T}_{\mu\nu}$ in accordance with expression (\ref{eq:T_tilde_decomp}) were determined up to a constant terms $\tilde{T}_{\mu\nu}^{(\text{const})}$. Thus, expressions of the tensor $\delta\tilde{T}_{\mu\nu}$ in the limit $\delta\omega\to0$ are defined as follows
\begin{align}
&\delta\tilde{T}^{(nw)}_{\mu\nu}\rightarrow\delta\tilde{T}^{(nw)}_{\mu\nu}+\tilde{T}_{\mu\nu}^{(\text{const})}.
\end{align}

However, we note that the main result for the study of the gravitational-wave perturbations of the space-time metric is that solutions \eqref{eq:h_00}--\eqref{eq:h_22_33} are regular ones in the limit $\delta\omega\to0$. We also note that for $n=1$ all non-wave terms (\ref{eq:energynw})--(\ref{eq:energynw5}) are equal to zero.

From equations \eqref{eq:deltaT_00}--\eqref{eq:deltaT_33}, \eqref{eq:h_00}--\eqref{eq:h_22_33} and \eqref{eq:R_0101}--\eqref{eq:R_1212} it follows that in the vacuum limit $n\to1$ all components of the tensor $\delta\tilde{T}_{\mu\nu}$, the metric perturbations and the Riemann tensor are equal to zero
\begin{align}
&\delta\tilde{T}_{\mu\nu}(n\to1)=0,\\
&h_{\mu\nu}(n\to1)=0,\\
&R_{\sigma\alpha\rho\beta}(n\to1)=0.
\end{align}

Also, solutions of equations (\ref{eq:wave_eq_delta}) in vacuum for
\begin{align}
\label{eq:deltaT_00vac}
&\delta\tilde{T}^{(vac)}_{00} = \frac{E_{0}^{2}}{4\pi} \cos\left( \delta\omega\left(t-\frac{x}{c}\right) \right),\\
\label{eq:deltaT_01vac}
&\delta\tilde{T}^{(vac)}_{01} = -\frac{E_{0}^{2}}{4\pi} \cos\left( \delta\omega\left(t-\frac{x}{c}\right) \right),\\
\label{eq:deltaT_11vac}
&\delta\tilde{T}^{(vac)}_{11} = \frac{E_{0}^{2}}{4\pi} \cos\left( \delta\omega\left(t-\frac{x}{c}\right) \right),\\
\label{eq:deltaT_22vac}
&\delta\tilde{T}^{(vac)}_{22} = \delta\tilde{T}_{33}=0,
\end{align}
can be written as follows
\begin{align}
&h^{(vac)}_{00} = -\frac{2 E_{0}^{2} G x \sin\left( \delta\omega\left(t-\frac{x}{c}\right) \right)}{\delta\omega c^{3}},
\label{eq:h_00_limit} \\
&h^{(vac)}_{01}  = \frac{2 E_{0}^{2} G x \sin\left( \delta\omega\left(t-\frac{x}{c}\right) \right)}{\delta\omega c^{3}},
\label{eq:h_01_limit} \\
&h^{(vac)}_{11} = -\frac{2 E_{0}^{2} G x \sin\left( \delta\omega\left(t-\frac{x}{c}\right) \right)}{\delta\omega c^{3}},
\label{eq:h_11_limit} \\
&h^{(vac)}_{22} = h^{(vac)}_{33} = 0,
\label{eq:h_22_33_limit}
\end{align}

From expression (\ref{eq:Riemann_weak}) we obtain that all components of the Riemann tensor for solutions (\ref{eq:h_00_limit})--(\ref{eq:h_22_33_limit}) are equal to zero $R^{(vac)}_{\sigma\mu\rho\nu}=0$.

We note that the sum of all longitudinal components is
\begin{align}
&h^{(vac)}_{00}+2h^{(vac)}_{01}+h^{(vac)}_{11}=0.
\label{z}
\end{align}

Also, the scalar curvature
\begin{align}
&R=\eta^{\mu\nu}R_{\mu\nu}=-\frac{1}{2}\eta^{\mu\nu}\eta^{\sigma\alpha} \partial_{\sigma} \partial_{\alpha} h_{\mu\nu} \nonumber\\
&=\frac{8\pi G}{c^{4}} \eta^{\mu\nu}\delta\tilde{T}_{\mu\nu}=\frac{8\pi G}{c^{4}}\delta\tilde{T}\nonumber\\
&=-\frac{6G E_{0}^{2}}{c^{4}}(n^{2}-1)\cos\left( \delta\omega\left(t-\frac{nx}{c}\right)\right),
\label{R}
\end{align}
vanishes in a vacuum $n=1$, and it is non-zero in a medium $n\neq1$.

Corresponding Kretschmann scalar~\cite{Cherubini:2002gen} can be obtained from the components of the Riemann tensor (\ref{eq:R_0101})--(\ref{eq:R_1212}) as $K=R_{\sigma\alpha\rho\beta}R^{\sigma\alpha\rho\beta}$.

Thus, based on the definition of the Kretschmann scalar we get following expression
\begin{align}
&K=R_{\sigma\alpha\rho\beta}R^{\sigma\alpha\rho\beta}\nonumber \\
& = \frac{4G^{2}E_{0}^{4}}{c^{8}} \Bigg\{
\big(3n^{4}+2n^{2}+3\big) \cos^{2}\!\left(\delta\omega\Big(t-\frac{nx}{c}\Big)\right) \nonumber \\
&- 16n \cos\!\left(\delta\omega\Big(t-\frac{nx}{c}\Big)\right)\cos\!\left(\delta\omega\Big(t-\frac{x}{c}\Big)\right)  \nonumber \\
&+ 4\left[2 + \left(\frac{n-1}{n+1}\right)^{2}\right] \cos^{2}\!\left(\delta\omega\Big(t-\frac{x}{c}\Big)\right)
 \Bigg\}.
\label{Kretschmann}
\end{align}

Under condition $n=1$ the Kretschmann scalar $K=0$, and in general case $n\neq1$ from (\ref{Kretschmann}) it follows that $K\neq0$.

As one can see, from expressions (\ref{R})--(\ref{Kretschmann}) it follows that the scalar curvature and Kretschmann scalar are finite throughout space-time. Thus, solutions (\ref{eq:h_00})--(\ref{eq:h_22_33}) are regular throughout space-time as well.
We also note that, unlike scalar curvature (\ref{R}), Kretschmann scalar (\ref{Kretschmann}) contains contributions from two types of gravitational waves with the different phase velocities $v^{(A)}_{gw}=c/n$ and $v^{(B)}_{gw}=c$.

Thus, vacuum solutions (\ref{eq:deltaT_00vac})--(\ref{eq:h_22_33_limit}) don't correspond to the real perturbations of the space-time metric and can be eliminated by additional gauge transformations. Therefore, in this case, the difference between the perturbations of the space-time metric in the material media and in a vacuum is equivalent to the perturbations of the space-time metric in the media themselves.

In the material media with $n \ne 1$, solutions \eqref{eq:h_00}--\eqref{eq:h_22_33} describe longitudinal and transverse components of the gravitational waves implying non-zero and finite invariants (\ref{R})--(\ref{Kretschmann}) and corresponding non-zero components (\ref{eq:R_0101})--(\ref{eq:R_1212}) of the Riemann tensor. Therefore, coupled gravitational waves described by solutions \eqref{eq:h_00}--\eqref{eq:h_22_33} in a material media can influence the test masses.

\section{Coupled gravitational waves in the interferometers}

Now, we consider the electromagnetic field with components \eqref{eq:Ey}--\eqref{eq:Hz} in the Michelson interferometer filed with a magnetized cold plasma or a rarefied neutral gas. Let us consider in more detail the components $R_{i0j0}$, which affect the distance between test masses (mirrors).

Based on the equation of geodesic deviation in the Newtonian approximation~\cite{Maggiore:2018sht}
\begin{eqnarray}
&&a^{i} = \frac{d^{2} \xi^{i}}{dt^{2}} \approx -c^{2} R^{i}{}_{0j0} \xi^{j},
\label{eq:geodesic_dev}
\end{eqnarray}
it can be concluded that the coupled gravitational waves induce acceleration of test masses by a tidal forces.

Solutions (\ref{eq:h_00})--(\ref{eq:h_11}) imply that two types of coupled gravitational waves with different phase velocities propagate along with the electromagnetic wave.

Thus, we can firstly consider the longitude coupled gravitational waves with the phase velocity $v^{(A)}_{gw}=c/n$ and following components
\begin{align}
h^{(A)}_{00} = \frac{2 G E_{0}^{2} (n^{2}-3)}{c^{2} (\delta\omega)^{2} (n^{2}-1)}
          \cos\left( \delta\omega\left(t-\frac{nx}{c}\right) \right),
\label{eq:h_00A}
\end{align}
\begin{align}
h^{(A)}_{01} = \frac{4 n G E_{0}^{2}}{c^{2} (\delta\omega)^{2} (n^{2}-1)}
                   \cos\left( \delta\omega\left(t-\frac{nx}{c}\right) \right),
\label{eq:h_01A}
\end{align}
\begin{align}
h^{(A)}_{11} = -\frac{2 G E_{0}^{2} (3n^{2}-1)}{c^{2} (\delta\omega)^{2} (n^{2}-1)}
          \cos\left( \delta\omega\left(t-\frac{nx}{c}\right) \right),
\label{eq:h_11A}
\end{align}

Corresponding component of the Riemann tensor can be written as follows
\begin{align}
\nonumber
&R_{1010(A)}= \frac{G E_{0}^{2} (n^{2}-1)}{c^{4}}
      \cos\left( \delta\omega\left(t-\frac{nx}{c}\right) \right) \\
&= \frac{1}{4}(n+1)^{2}\partial_{0}\partial_{0}h^{1}_{A}=\frac{1}{4c^{2}}(n+1)^{2}\ddot{h}^{1}_{A},
\label{eq:R_0101A}
\end{align}
where
\begin{align}
&h^{1}_{A}\equiv h^{(A)}_{00}+2h^{(A)}_{01}+h^{(A)}_{11},
\label{HA}
\end{align}
and $\partial_{0}\partial_{0}h^{1}_{A}=\frac{1}{c^{2}}\ddot{h}^{1}_{A}$ in the proper detector frame for the fixed mirror's positions.

Considering a small shift in the distance between the mirrors of interferometer at the $x$--axis direction
\begin{eqnarray}
&&\xi^{1}(t)=L_{x}+\delta x(t)=L_{x}+\Delta L_{x},
\label{shiftA}
\end{eqnarray}
and taking into account that
\begin{eqnarray}
&&-R^{i}{}_{0j0}=R_{i0j0},~~(i,j=1,2,3),
\label{RiemannM}
\end{eqnarray}
from equations (\ref{eq:geodesic_dev}) and (\ref{eq:R_0101A}) we obtain
\begin{eqnarray}
&&\frac{d^{2}}{dt^{2}}\left(\delta x(t)\right)\approx\frac{1}{4}(n+1)^{2}L_{x}\ddot{h}^{1}_{A}.
\label{shift2A}
\end{eqnarray}

After substitution (\ref{eq:h_00A})--(\ref{HA}) into (\ref{shift2A}) we get following relative shift of the mirrors inspired by the coupled longitude gravitational waves with the phase velocity $v^{(A)}_{gw}=c/n$ at the $x$--axis direction
\begin{align}
\nonumber
&\left(\frac{\Delta L_{x}}{L_{x}}\right)_{(A)}\approx \frac{(n+1)^{2}}{4}\left(h^{(A)}_{00}+2h^{(A)}_{01}+h^{(A)}_{11}\right)\\
&=h^{1}_{0A}\cos\left( \delta\omega\left(t-\frac{nx}{c}\right) \right),
\label{shiftA2}
\end{align}
with amplitude
\begin{eqnarray}
\nonumber
&&h^{1}_{0A}=-\frac{G E_{0}^{2}}{c^{2} (\delta\omega)^{2}}\left(\frac{n-1}{n+1}\right)(n+1)^{2}\\
&&=-\frac{G E_{0}^{2}(n^{2}-1)}{c^{2} (\delta\omega)^{2}}.
\label{shiftA3}
\end{eqnarray}

The transverse coupled gravitational waves with the phase velocity $v^{(A)}_{gw}=c/n$ have the following components
\begin{align}
&h^{(A)}_{22} = h^{(A)}_{33} = -\frac{2 G E_{0}^{2}}{c^{2} (\delta\omega)^{2}}
                   \cos\left( \delta\omega\left(t-\frac{nx}{c}\right) \right).
\label{eq:h_22_33A}
\end{align}

Corresponding components of the Riemann tensor are
\begin{align}
&R_{2020(A)} =R_{3030(A)}= -\frac{G E_{0}^{2}}{c^{4}} \cos\left( \delta\omega\left(t-\frac{nx}{c}\right) \right)\nonumber\\
&=-\frac{1}{2} \partial_{0} \partial_{0} h^{(A)}_{22}= -\frac{1}{2} \partial_{0} \partial_{0} h^{(A)}_{33}.
\label{HTA}
\end{align}

From (\ref{eq:geodesic_dev}) and (\ref{eq:h_22_33A})--(\ref{HTA}) we obtain following relative shifts of the mirrors in $y$ or $z$ direction
\begin{eqnarray}
&&\left(\frac{\Delta L_{y,z}}{L_{y,z}}\right)_{(A)}\approx h^{2,3}_{0A}\cos\left( \delta\omega\left(t-\frac{nx}{c}\right) \right),
\label{shiftAT2}
\end{eqnarray}
where
\begin{eqnarray}
&&h^{2,3}_{0A}=\frac{G E_{0}^{2}}{c^{2} (\delta\omega)^{2}}.
\label{shiftAT3}
\end{eqnarray}

Thus, the phase shift in the interferometer induced by the coupled gravitational waves with phase velocity $v^{(A)}_{gw}=c/n$ is
\begin{eqnarray}
\nonumber
&&\Delta\phi_{A}=\frac{4\pi n}{\lambda_{0}}\left[\Delta L_{x}-\Delta L_{y,z}\right]\\
&&\approx-\frac{2G E_{0}^{2}\,\omega_{0}n}{c^{3}(\delta\omega)^{2}}\left[(n^2-1)L_{x}+L_{y,z}\right],
\label{MICH1}
\end{eqnarray}
where $\lambda_{0}=\frac{2\pi c}{\omega_{0}}$, and the coefficient $2$ in the optical path difference characterizes the passage of the electromagnetic wave back and forth.

Now, let us consider the longitude coupled gravitational waves with the phase velocity $v^{(B)}_{gw}=c$ and following components
\begin{align}
\label{eq:h_00B}
&h^{(B)}_{00} =  - \frac{2 G E_{0}^{2} (n^{2}-3)}{c^{2} (\delta\omega)^{2} (n^{2}-1)}
          \cos\left( \delta\omega\left(t-\frac{x}{c}\right) \right)\nonumber\\
&+\frac{2 E_{0}^{2} G x}{\delta\omega c^{3}}\sin\left( \delta\omega\left(t-\frac{x}{c}\right) \right),
\end{align}
\begin{align}
\label{eq:h_01B}
&h_{01}^{(B)} =  - \frac{4 n G E_{0}^{2}}{c^{2} (\delta\omega)^{2} (n^{2}-1)}
                   \cos\left( \delta\omega\left(t-\frac{x}{c}\right) \right)\nonumber\\
&-\frac{2 E_{0}^{2} G x}{\delta\omega c^{3}}\sin\left( \delta\omega\left(t-\frac{x}{c}\right) \right),
\end{align}
\begin{align}
\label{eq:h_11B}
&h_{11}^{(B)} = \frac{2 G E_{0}^{2} (3n^{2}-1)}{c^{2} (\delta\omega)^{2} (n^{2}-1)}
          \cos\left( \delta\omega\left(t-\frac{x}{c}\right) \right)\nonumber\\
&+\frac{2 E_{0}^{2} G x}{\delta\omega c^{3}}\sin\left( \delta\omega\left(t-\frac{x}{c}\right) \right)
\end{align}

Corresponding component of the Riemann tensor is
\begin{align}
\nonumber
&R_{1010(B)}= \frac{2 G E_{0}^{2} (n-1)}{(n+1) c^{4}}
      \cos\left( \delta\omega\left(t-\frac{x}{c}\right) \right) \\
&= -\frac{1}{2}\partial_{0}\partial_{0}h^{1}_{B}=-\frac{1}{2}\ddot{h}^{1}_{B},
\label{eq:R_0101B}
\end{align}
where
\begin{align}
&h^{1}_{B}\equiv h^{(B)}_{00}+2h^{(B)}_{01}+h^{(B)}_{11}.
\label{HB}
\end{align}

From equation (\ref{eq:geodesic_dev}) and expressions (\ref{eq:h_00B})--(\ref{HB}) we obtain following relative shift of the mirrors inspired by the coupled longitude gravitational waves with the phase velocity $v^{(B)}_{gw}=c$ at the $x$--axis direction
\begin{eqnarray}
&&\left(\frac{\Delta L_{x}}{L_{x}}\right)_{(B)}\approx h^{1}_{0B}\cos\left( \delta\omega\left(t-\frac{x}{c}\right) \right),
\label{shiftB2}
\end{eqnarray}
with amplitude
\begin{eqnarray}
&&h^{1}_{0B}=-\frac{2G E_{0}^{2}}{c^{2} (\delta\omega)^{2}}\left(\frac{n-1}{n+1}\right).
\label{shiftB3}
\end{eqnarray}

The transverse coupled gravitational waves with the phase velocity $v^{(B)}_{gw}=c$ have the following components
\begin{align}
&h^{(B)}_{22} = h^{(B)}_{33} = \frac{2 G E_{0}^{2}}{c^{2} (\delta\omega)^{2}}
                   \cos\left( \delta\omega\left(t-\frac{x}{c}\right) \right).
\label{eq:h_22_33B}
\end{align}

Corresponding components of the Riemann tensor are
\begin{align}
&R_{2020(B)} =R_{3030(B)}= \frac{G E_{0}^{2}}{c^{4}} \cos\left(\delta\omega\left(t-\frac{x}{c}\right) \right)\nonumber\\
&=-\frac{1}{2} \partial_{0} \partial_{0} h^{(B)}_{22}= -\frac{1}{2} \partial_{0} \partial_{0} h^{(B)}_{33}.
\label{HTB}
\end{align}

From (\ref{eq:geodesic_dev}) and (\ref{eq:h_22_33B})--(\ref{HTB}) we obtain following relative shifts of the mirrors in $y$ or $z$ direction
\begin{eqnarray}
&&\left(\frac{\Delta L_{y,z}}{L_{y,z}}\right)_{(B)}\approx h^{2,3}_{0B}\cos\left( \delta\omega\left(t-\frac{x}{c}\right) \right),
\label{shiftBT2}
\end{eqnarray}
where
\begin{eqnarray}
&&h^{2,3}_{0B}=-\frac{G E_{0}^{2}}{c^{2} (\delta\omega)^{2}}=-h^{2,3}_{0A}.
\label{shiftBT3}
\end{eqnarray}

Corresponding phase shift in the interferometer induced by the coupled gravitational waves with phase velocity $v^{(B)}_{gw}=c$ is
\begin{align}
&\Delta\phi_{B}
\approx-\frac{G E_{0}^{2}\,\omega_{0}n}{c^{3}(\delta\omega)^{2}}\left[2\left(\frac{n-1}{n+1}\right)L_{x}-L_{y,z}\right],
\label{MICH2}
\end{align}

Because the propagation velocities of the gravitational waves are different, phase shifts $\Delta\phi_{A}$ and $\Delta\phi_{B}$ must generally be considered separately. However, we will consider two limiting cases.

For the case $n\gg1$ the resulting phase shift for $L_{x}\equiv L$ can be written as follows
\begin{align}
&\Delta\phi\approx-\frac{2G E_{0}^{2}\,\omega_{0}n^3L}{c^{3}(\delta\omega)^{2}}=\frac{4\pi n}{\lambda_{0}}h_{0}(n\gg1)L.
\label{RES1}
\end{align}

This phase shift is induced by the coupled gravitational waves with amplitude
\begin{align}
&|h_{0}(n\gg1)|\approx\frac{G E_{0}^{2}n^{2}}{c^{2} (\delta\omega)^{2}}=\frac{8\pi G I_{em} n^{2}}{c^{3} (\delta\omega)^{2}},
\label{RES2A}
\end{align}
where $I_{em}=\frac{c}{8\pi}E_{0}^{2}$ is the intensity of electromagnetic wave in a vacuum.

For the case $n=1+\delta n$, where $\delta n\ll1$, the resulting phase shift is
\begin{align}
&\Delta\phi\approx\Delta\phi_{A}+\Delta\phi_{B}\approx-\frac{3G E_{0}^{2}\,\omega_{0}}{c^{3}(\delta\omega)^{2}}\delta n L
\nonumber\\
&=\frac{4\pi}{\lambda_{0}}h_{0}(\delta n\ll1)L.
\label{MICHSM}
\end{align}

This phase shift is induced by the coupled gravitational waves with amplitude
\begin{align}
&|h_{0}(\delta n\ll1)|\approx\frac{3G E_{0}^{2}\delta n}{2c^{2} (\delta\omega)^{2}}=\frac{12\pi G I_{em}\delta n}{c^{3} (\delta\omega)^{2}}.
\label{RES2B}
\end{align}

The amplification factor of the amplitude of gravitational waves for the case $n\gg1$ is
\begin{align}
&\eta=\frac{|h_{0}(n\gg1)|}{|h_{0}(\delta n\ll1)|}\approx\frac{n^{2}}{\delta n}.
\label{AMP}
\end{align}

In both these cases, the phase shift is determined only by the longitudinal components of the gravitational waves coupled with the electromagnetic wave.

Thus, in case of the difference of electromagnetic radiation from monochromatic one $\delta\omega \ne 0$ and deviations of the refractive index from unity $n \ne 1$, longitudinal gravitational waves coupled with the electromagnetic waves induce the phase shift in interferometric gravitational wave detectors similar to transverse free gravitational waves from external sources.
A characteristic difference between coupled longitudinal gravitational waves and external transverse gravitational waves is that their frequency is determined by the deviation $\delta\omega$ from monochromatic electromagnetic radiation, which is a controlled parameter.
Nevertheless, this effect is negligible for a LIGO-type interferometers with a refractive index $n\simeq1$~\cite{AdvLIGO:2021oxw}.

However, we can estimate the possible amplitude of the coupled gravitational waves with a refractive index different from unity.

As an example, let us consider laser radiation with very high intensity in a rarefied neutral gas.
For the following parameters:
\begin{align}
&I_{\text{em}}\sim10^{6}\ \text{W/cm}^{2}=10^{10}\ \text{W/m}^{2},\\
&\delta\omega\sim100\ \text{Hz},~~~\delta n\sim 10^{-2},
\label{PAR}
\end{align}
from expression (\ref{RES2B}) we obtain $h_{0}\sim10^{-30}$.

On the other hand, considering a cold magnetized plasma with refractive index $n\sim 10^{3}$~\cite{Shklyar_2025} instead of a rarefied neutral gas from expression (\ref{AMP}) we obtain $h_{0}\sim10^{-22}$ for the same intensity and frequency.
This value is comparable to the amplitude of transverse free gravitational waves with a frequency $\omega_{gw}\sim100$ Hz from external astrophysical sources.
The main problem, in this case, is the separation of the gravitational wave signal from the influence of the processes accompanying the interaction between electromagnetic radiation and the cold magnetized plasma on the change in the intensity of electromagnetic radiation on the photodetector.

For the promising cosmic detectors LISA~\cite{LISA:2022yao,LISAConsortiumWaveformWorkingGroup:2023arg} the inhomogeneous and anisotropic refractive index of the solar plasma corresponding to the arising coupled gravitational waves. In principle, it is possible to estimate the amplitude of longitudinal coupled gravitational waves by expression (\ref{RES2B}) for the certain average value of the deviation of the refractive index from unity along the propagation of electromagnetic radiation $\langle \delta n\rangle\ll1$.
Despite the small amplitude of the coupled gravitational waves, the long arm's length $L\sim10^{6}$ km increases the possible influence of this type of the gravitational waves on the phase shift compared to a LIGO-type detectors.

\section{Conclusion}

This paper examined the properties of gravitational waves coupled with electromagnetic waves in material media. As an amplifying factor for the amplitude of gravitational waves, we considered the deviation of electromagnetic radiation from a monochromatic one with difference frequency $\delta\omega$. Also, a factor directly influencing the invariants of perturbations of the space-time metric is the refractive index of the medium. In particular, the scalar curvature of the space-time metric in the field of a gravitational wave coupled with an electromagnetic wave in a medium depends on the refractive index as $R\sim(n^{2}-1)$. This result means that in a vacuum there are no coupled gravitational waves with a difference frequency $\delta\omega$. On the other hand, in material media with a high refractive index $n\gg1$, deviations from zero scalar curvature increase by a factor proportional to $n^{2}$, and corresponding deviations of the Kretschmann scalar from zero value increase by a factor proportional to $n^{4}$. We also note that non-zero and finite values of the scalar curvature (Ricci scalar) and the Kretschmann scalar mean that the gravitational waves solutions under consideration correspond to real perturbations of the space-time metric which cannot be eliminated by the coordinate transformations.

It has also been shown that longitudinal gravitational waves coupled with electromagnetic waves in the medium have an effect on test masses similar to transverse gravitational waves from external sources. The fundamental problem of interferometric registration of such gravitational waves is the appearance of additional noise due to the interaction of the electromagnetic wave and the medium.
Despite this fact, it can be noted that the amplitude of longitudinal gravitational waves coupled with electromagnetic waves in a medium can significantly exceed the amplitude of transverse gravitational waves induced by powerful pulses of laser radiation~\cite{Kadlecova:2016fbx,Fomin:2020ovn,Lageyre:2021cxe,Atonga:2023psc}.


Thus, there is some interest in a generalized studying this type of gravitational waves in the vicinity of compact astrophysical objects or in the early universe, taking into account the proposed solutions as a limiting case.

\end{document}